\documentclass[10pt]{article}

\usepackage{amsmath}
\usepackage{amssymb}

\usepackage{graphicx}

\usepackage{cite}

\usepackage{color} 

\usepackage[labelfont=bf,labelsep=period,justification=raggedright]{caption}

\makeatletter
\renewcommand{\@biblabel}[1]{\quad#1.}
\makeatother

\date{}

\begin{document}

\begin{flushleft}
{\Large
\textbf{Noise-aided Logic in an Electronic Analog of Synthetic Genetic
  Networks}
}
\\
Edward H. Hellen$^{1,\ast}$, 
Syamal K. Dana$^{2}$,
J\"{u}rgen Kurths$^{3}$,
Elizabeth Kehler$^{1}$, 
Sudeshna Sinha$^{4}$
\\
\bf{1} Department of Physics and Astronomy, University of North Carolina Greensboro, 
  Greensboro, NC 27402, USA
\\
\bf{2} CSIR-Indian Institute of Chemical Biology, Kolkata 700032, India
\\
\bf{3} Potsdam Institute for Climate Impact Research,
    Telegrafenberg A31 14473 Potsdam 14473 Potsdam, Germany
\\
\bf{4} Indian Institute of Science Education and Research
  (IISER) Mohali, SAS Nagar, Sector 81, Mohali 140 306, Punjab, India
\\
$\ast$ E-mail: ehhellen@uncg.edu
\end{flushleft}

\section*{Abstract}
We report the experimental verification of noise-enhanced logic
  behaviour in an electronic analog of a synthetic genetic network,
  composed of two repressors and two constitutive promoters.  We
  observe good agreement between circuit measurements and numerical
  prediction, with the circuit allowing for robust logic operations in
  an optimal window of noise. Namely, the input-output characteristics
  of a logic gate is reproduced faithfully under moderate noise, which
  is a manifestation of the phenomenon known as {\em Logical
    Stochastic Resonance}. The two dynamical variables
  in the system yield complementary logic behaviour simultaneously.  The system is easily morphed from AND/NAND to OR/NOR logic. 


\section*{Introduction}
Realization of logic functions in different physical systems is one of
the key questions that commands widespread research interest in
science and engineering. Universal general-purpose computing devices
can be constructed entirely from NOR/NAND logic gates \cite{mano1,bartee}. It is
particularly interesting to investigate if systems of biological
relevance can also yield logic outputs consistent with the truth
tables of different logic functions (see Table \ref{table1}). Biological systems are capable of stochastic resonance\cite{Bulsara1991, Levin1996, Gammaitoni1998, Hanggi2002}, a process in which a small signal is amplified due to the presence of an appropriate level of noise, leading to the possibility of a biological system performing robust noise-aided logic operations in response to weak input signals. 

A new idea in this direction uses the interplay between noise and
nonlinearity constructively to enhance the robustness of logic
operations. Namely, in an optimal window of noise, the input-output
characteristics of a logic gate is reproduced faithfully. This
phenomenon is termed {\em Logical Stochastic Resonance} (LSR)
\cite{lsrprl,lsr1,lsr2,lsr3,lsr4,lsr5}. Specifically, in LSR we
consider the state of a nonlinear system when driven by input signals,
consisting of two randomly streaming square waves. It was observed
that the response of such a system shows a remarkable feature: in an
optimal band of noise, the output of the system, determined by its
state, is a logical combination of the two input signals in accordance
with the truth tables of fundamental logic operations. 

An important motivation for further studying LSR stems from an issue
that is receiving widespread attention currently. The
  number of transistors in an integrated circuit has approximately
  doubled every year in accordance with Moore's law. The rapid
  shrinking of computing platforms with smaller power supplies has
  brought with it problems of smaller noise margins and higher error
  rates. Namely, as computational devices and platforms continue to
shrink in size, we encounter fundamental noise that cannot be
suppressed or eliminated.  Hence an understanding of the cooperative
behavior between a device noise-floor and its nonlinearity plays an
increasingly crucial role in paving the way for smart
  computing devices.  In this direction, LSR indicates a way to turn
potentially performance degrading noise to assist the desired
operation. Further, it is of far reaching interest to obtain analogous
behaviour, not merely in human engineered physical systems, but also
in systems of chemical and biological relevance, in
  order to explore the information processing capacity of naturally
  occurring systems where noise is ubiquitous.

Since the idea of LSR was first introduced
  \cite{lsrprl}, several systems implementing and displaying LSR have
  been found. To begin with, the basic electronic realizations of
  simple bistable potentials were reported \cite{lsrprl,lsr1}.
  Subsequently, noise-aided reprogrammable logic gates have been
  implemented with noisy nanomechanical oscillators \cite{lsr5},
  chemical systems \cite{lsr5} and optical systems \cite{laser,
  vcsels}.

Most recently, in the context of biological systems,
  theoretical ideas have been proposed \cite{Ando2011,gene2,Dari2011,Sharma2013} on the
  implementation of LSR in a synthetic genetic network \cite{hasty}.
  Now, in this work, we will provide {\em experimental realizations}
  of these ideas in an electronic analog of a noisy synthetic gene
  network.  Specifically then, we will investigate the possibility of
  obtaining reliable logic outputs, and {\em explicitly demonstrate
    the pivotal role of noise in the optimization of the logic
    performance in this circuit}.  Further, we will show that the system is easily changed from AND/NAND logic to OR/NOR demonstrating potential for re-programmability \cite{Ando2011,gene2}. Our results
  will thus provide verification and further understanding of noise aided
  logic in systems that are of considerable importance in biology.

Since understanding the intracellular processes in a network of
interacting biomolecules is difficult, an alternative approach has
been started recently \cite{elowitz1}, to design artificial genetic
networks to derive desired functional behaviors. One important early
design is a clock using three genes inhibiting each other in a cyclic
order \cite{elowitz2}. Taking into account the standard chemical
kinetics for expression, degradation and inhibition, a dynamical
system model was proposed where the repressor-protein concentrations
and mRNA concentrations were expressed as dynamical variables. Another
design is a synthetic genetic toggle-switch network \cite{gardner}
whose potential for noise-aided logic operation is investigated here.

The significance of using both numerical simulation and electronic
circuits to model a potential synthetic genetic network is two-fold.
Firstly, the numerical and circuit methods provide {\em different}
imperfect models of a potential biological system. Agreement between
these two models indicates robustness in the system and therefore
greater likelihood that the same behavior could be realized in the
proposed biological system.  The biological system is generally much
more difficult to construct, and therefore
investigating proposed networks in simpler systems is
  prudent. Secondly, modeling with stochastic differential equations
is nontrivial compared to ordinary differential equations, so that the
addition of experimental measurements from a physical system such as
an analog circuit provides valuable verification.  Thus the circuit is
an additional tool for investigating potentially interesting
biological networks in the presence of noise.
 
Here we use two repressors and constitutive promoters as our model
system for implementing logic functions.  First we describe the
synthetic gene network model below, and define what constitutes logic
inputs and logic outputs in this system. We then go on to present the
electronic analog of the system followed by a comparison of numerical
simulation and experimental measurement.

\section*{Methods}
We begin with a short description of the general principle of LSR.
Consider a general nonlinear dynamic system, given by

\begin{equation}
\frac{dx}{dt} = F(x) + I + D \eta (t)
\end{equation}
where $F(x)$ is a generic nonlinear function which has or nearly has
two distinct stable energy wells. $I$ is a low amplitude input
signal and $\eta (t)$ is an additive zero-mean Gaussian noise with unit
variance with $D$ being the noise strength.

We achieve a logical input-output correspondence with such a system
by encoding N inputs in N square waves. Specifically, for two logic
inputs, we drive the system with a low amplitude signal $I$, taken to be
the sum of two pulse trains: $I_1+I_2$, where $I_1$ and $I_2$ encode the two
logic inputs. Now the logic inputs can be either 0 or 1, giving rise
to 4 distinct logic input sets $(I_1,I_2)$: (0,0), (0,1), (1,0) and
(1,1). Since the input sets (0,1) and (1,0) give rise to the same $I$,
the input signal generated by adding two independent input signals is
a 3-level aperiodic waveform.

The output of the system is determined by its state. For instance, for
a bistable system with wells at $x= x_1$ and $x_2$, the output can be
considered a logical 1 if it is in the well at $x_1$, and logical 0 if it
is in $x_2$. If we consider the opposite assignment, namely logical 1 if
the state is in well $x_2$ and logical 0 if the state is in well $x_1$, we
obtain a complementary logic operation. Specifically we can have an
output determination threshold $x^*$, located near the barrier between
the wells, and the logical outputs are then simply given by the state
being greater than or less than $x^*$. It is possible that the input $I$ induces the appearance of the second energy well if it was not already there. 

The central result of LSR is as follows: for a given set of inputs
$(I_1,I_2)$, a logical output, in accordance with the truth tables of
the basic logic operations, is consistently obtained only in an
optimal window of noise. Namely, under very small or very large noise
the system does not yield reliable logic outputs, while in a band of
moderate noise it produces the desired output.

\subsection*{Synthetic Genetic Network Model and Logic Operation}
We consider the previously used variation \cite{Ando2011} of the genetic toggle switch model comprised of two genes inhibiting each other\cite{gardner}. The
concentrations of the two expressed proteins are $x$ and $y$, and
their rates of change are:
\begin{eqnarray}
\frac{dx}{dt^\prime}&=&\frac{\alpha_1}{1+y^{n}}-\beta_1 x + g_1 + D \eta (t^\prime) 
\label{gene1}\\
\frac{dy}{dt^\prime}&=&\frac{\alpha_2}{1+x^{n}}-\beta_2 y + g_2 + I_1 + I_2  + D \xi (t^\prime)
\label{gene2}
\end{eqnarray}
where $\beta_1$, $\beta_2$ are the rates of decay of each expressed
protein and $n$ is the Hill coefficient.  The $\alpha_1$, $\alpha_2$
describe the maximum expression rates in absence of inhibitor and they
are used here as tunable parameters. In the original model $g_1$ and
$g_2$ represent the basal synthesis rates of the promoters
\cite{wang}, however we use them as constant bias. The additive noise
has strength $D$ and $\eta$ and $\xi$ are chosen from unit variance
zero mean Gaussian distributions. Such an additive noise
source alters the background repressor production and represents the
inherent stochasticity of biochemical processes such as transcription and
translation, and the fluctuations in the concentration of a regulatory
protein. $I_1$ and $I_2$ are two {\em low
  amplitude} inputs providing independent parallel production pathways of repressor
  \textit{y}. The $t^\prime$ indicates dimensionless time.

\begin{figure}[!ht]
\begin{center}
\includegraphics[width=4in]{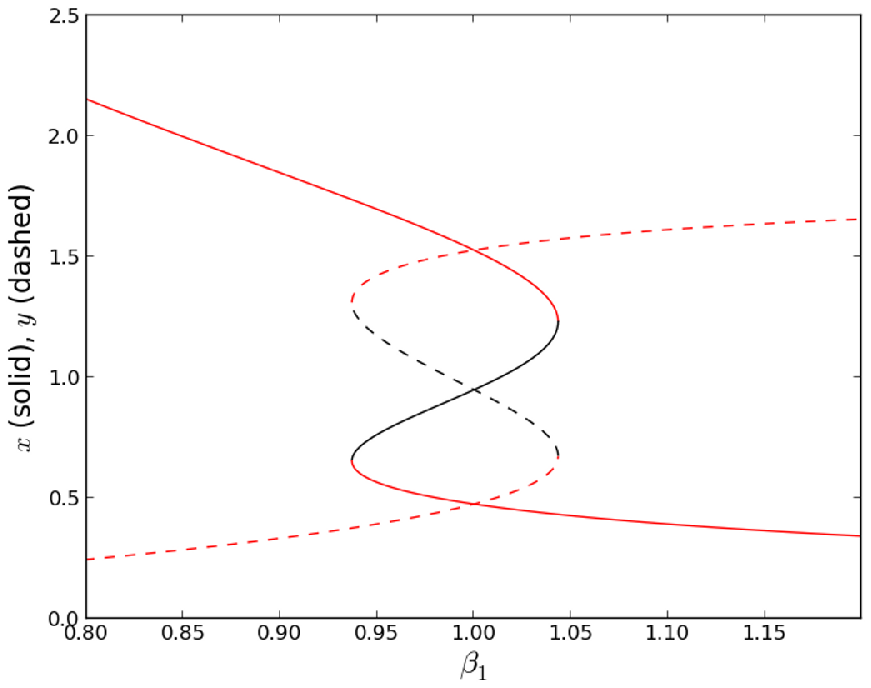}
\end{center}
\caption{{\bf Bifurcation diagram for state-variables $x$ (solid) and $y$ (dashed).} $x$ and $y$ are complementary outputs, when one is high the other is low. Red (black) indicate stable (unstable) fixed points. System is bistable for $0.937<\beta_1 <1.043$. For $\beta_1=0.9$, $x$ is high and $y$ is low. Calculated for Eqs.\ \eqref{gene1}-\eqref{gene2} with $n=2.4$, $\alpha_i=1.78$, $\beta_2=1$, $I_i=0$, $g_i=0$, and $D=0$. }
\label{fig:beta_bif}  
\end{figure}

The system above may have two stable configurations in the $xy$-plane: one
state has a high value of $x$ ($x_u$) and a low value of $y$ ($y_l$); the
other state has a low value of $x$ ($x_l$) and a high value of $y$
($y_u$). That is, the two dimensional potential underlying this system
has two wells, $(x_u, y_l)$ and $(x_l, y_u)$, in the $xy$-plane. Varying
the parameters changes the depth and position of these wells, and also determines whether there are one or two wells.  For example, Fig.\ \ref{fig:beta_bif} shows that for the case $(g_i,I_i,D)=0$ the system in Eqs.\ \eqref{gene1}-\eqref{gene2} is bistable and therefore has two stable wells only when $\beta_1$ is close to 1. 

{\bf Encoding Inputs}: Here the low amplitude input signal is $I = I_1
+I_2$, with $I_1/I_2$ equal to $I_{ON}$ ($I_{ON} > 0$) if the logic input
is $1$, and $I_1/I_2$ being $0$ if the logic input is $0$. So we have:

\noindent
(i)  $I_1 + I_2 = 0$ corresponds to logic input set $(0,0)$\\
(ii) $I_1 + I_2 =  I_{ON}$ corresponds to logic input sets $(0,1)$/$(1,0)$\\
(iii) $I_1 + I_2 = 2 I_{ON}$ corresponds to logic input set $(1,1)$\

{\bf Output}: The outputs of the system are determined by the level of the dynamical
variables $x(t)$ and $y(t)$. For instance the output can be considered
a logical $1$ if the state is at the high level, and logical $0$ if it
is at the lower level. That is:

\noindent
(i) If $x \ < \ x^*$, then Logic Output is $0$\\
(ii) If $x \ > \ x^*$, then Logic Output is $1$\

Here $x^*$ is the {\em output determination threshold} that lies
between the two states, e.g., at the position of the barrier
between the wells. The results presented here are not sensitive to the
specific value of $x^*$.

Specifically, in this work, we consider the logic output to be $1$
when the state is close to the upper well, and $0$ when the state is
close to the lower well, for both $x$ and $y$ variables.  So when the
system switches wells, the output is ``flipped'' or ``toggled''.

The model in Eqs.\ \eqref{gene1}-\eqref{gene2} is based on the synthetic genetic toggle switch previously expressed in \textit{E. coli} \cite{gardner}.  Parameter values used in \cite{gardner} correspond here to $\alpha_i = (15.6, 156)$, $n=(1, 2.5)$, and $\beta_i=1$ in Eqs.\ \eqref{gene1}-\eqref{gene2}. By comparison, here we use $\alpha_i = 1.78$, $n=2.4$, $\beta_1 =0.9$, and $\beta_2=1$.  The bifurcation diagram in Fig.\ \ref{fig:beta_bif} indicates that these parameter values, along with $(g_i,I_i,D)=0$, result in a system with a single stable well at $x\approx 1.8, y\approx 0.35$. A non-zero input $I$ can then "shift" the bifurcation diagram of Fig.\ \ref{fig:beta_bif} so that there is a stable state with low-$x$, high-$y$ at $\beta_1=0.9$. 

\subsection*{Circuit Realization}
\begin{figure}[!ht]
\begin{center}
\includegraphics[width=4in]{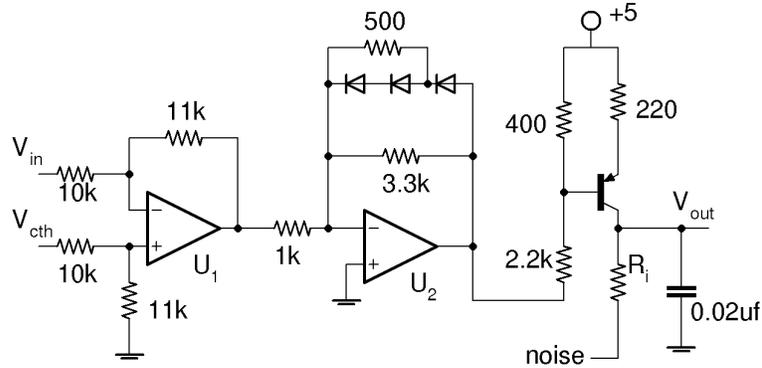}
\end{center}
\caption{{\bf Circuit for single gene.} Inhibitory input at
  $V_{in}$. Expressed protein concentration is represented by
  $V_{out}$. $R_i$ = 470 $\Omega$ for gene-\textit{y}, 520 $\Omega$
  for gene-\textit{x}. Dual op-amp is LF412 supplied by +/-5 V. The
  \textit{pnp} transistor is 2N3906. The input noise has a mean of 0 V
  (gnd) and controllable standard deviation.}
\label{fig:gene}  
\end{figure}

\begin{figure}[!ht]
\begin{center}
\includegraphics[width=4in]{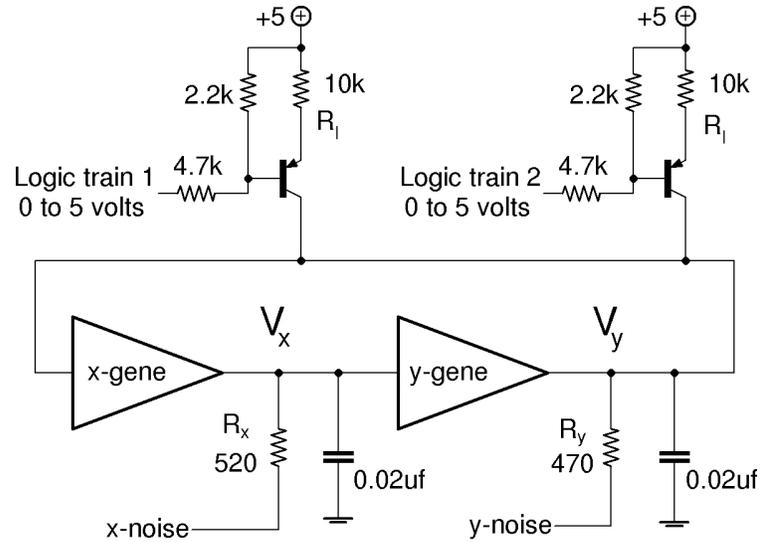}
\end{center}
\caption{{\bf Circuit for synthetic genetic
  network.} Encoding inputs are $0$ to $5$ V pulse trains creating transistor currents of $0.1$ and $0$ mA, respectively, for $R_I=10k\Omega$. The \textit{x}
  and \textit{y} gene circuits are shown in Fig.\ \ref{fig:gene}. Each noise input
  is connected to its own noise circuit shown in Fig.\ \ref{fig:noise_circuit}.}
\label{fig:lsr_network}  
\end{figure}

The circuit of a single inhibitory gene
\cite{ed,ed2} is shown in Fig.\ \ref{fig:gene}. The transistor current represents the rate of gene expression
and the voltage $V_{out}$ represents the concentration of expressed
protein.  $V_{in}$ represents the concentration of repressor, and the $V_{cth}$ adjusts the affinity of the repressor binding to the gene's DNA. 
The Hill function inhibition in Eqs.\ \eqref{gene1}-\eqref{gene2} is accounted for
by the dependence of the transistor current on repressor concentration voltage $V_{in}$. 
The synthetic genetic network shown in Fig.\ \ref{fig:lsr_network} is
comprised of two individual gene circuits connected in a loop, each
inhibiting the other.  For the model in Eqs.\ \eqref{gene1}-\eqref{gene2}, the encoding
inputs $I_1$ and $I_2$ add to production of $y$ which is accounted for in
Fig.\ \ref{fig:lsr_network} by the two logic-driven transistors
sourcing current to $V_{y}$.  Initially parameters $g_1$ and $g_2$ are taken to be zero.  

\begin{figure}[!ht]
\begin{center}
\includegraphics[width=4in]{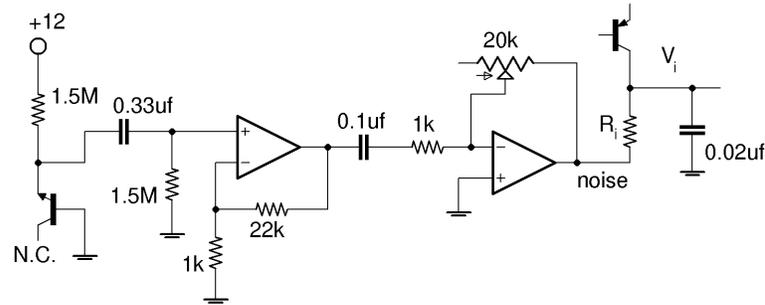}
\end{center}
\caption{{\bf Noise circuit and its connection to
  resistor of gene circuit.} Source of noise is the reverse-biased base-emitter junction of the 2N3904 \textit{npn} transistor on left. OPA2228 dual op-amps supplied from +/-12 V
  regulators. OPA2228 has gain-bandwidth product of 33 MHz.}
\label{fig:noise_circuit}  
\end{figure}

The circuit equations are obtained by applying Kirchoff's laws to $V_x$ and $V_y$, the voltages across the capacitors in Fig.\ \ref{fig:lsr_network} \cite{ed,ed2}. Multiplying both equations by $R_y$ results in equations for $V_x$ and $V_y$;
\begin{eqnarray}
R_yC\frac{dV_x}{dt}&=& -\frac{R_y}{R_x}V_x+\frac{R_y}{R_x}V_{x\_ noise}+R_yi_t
\label{circ_eq1}\\ 
R_yC\frac{dV_y}{dt}&=& -V_y+V_{y\_ noise}+R_yi_t+R_yi_1+R_yi_2
\label{circ_eq2}
\end{eqnarray}
where $i_t$ are the Fig.\ \ref{fig:gene} transistor currents for each gene, and $i_1$ and $i_2$ are currents from the logic train transistors in Fig.\ \ref{fig:lsr_network}. A noise generation circuit shown in Fig.\ \ref{fig:noise_circuit} based on
breakdown of a reverse biased base-emitter junction produces noise $V_{noise}$
with zero mean and variable amplitude.  We use a well regulated supply
for the noise circuit to avoid adding AC signals from the building's electrical system into the
noise. Two of these noise circuits are used to supply noisy voltages to each individual gene at locations indicated in Fig.\ \ref{fig:lsr_network}. 

The connections between model parameters ($\alpha_i, \beta_i, n, I_i, g_i,D$) and circuit parameters are presented in this section using relevant numerical values, with derivations of these connections given in the next section. Readers may go directly to \textit{Results and Discussion} without loss of continuity. The connections are found by relating circuit Eqs.\ \eqref{circ_eq1}-\eqref{circ_eq2} to the model Eqs.\ \eqref{gene1}-\eqref{gene2} and by adjusting the dependence of the transistor current $i_t$ on $V_{in}$ in Fig.\ \ref{fig:gene} to match the Hill function inhibition. The dimensionless state-variables ($x, y$) in Eqs.\ \eqref{gene1}-\eqref{gene2} are related to voltages ${V_x}$ and ${V_y}$ by:
\begin{equation*}
x = \frac{V_x}{V_{th}}, y = \frac{V_y}{V_{th}}
\end{equation*}  
where $V_{th}$ corresponds to the repressor's half-maximal inhibition binding constant $K_i$. The maximal expression rate is
\begin{equation}
\alpha_1=\alpha_2=\frac{i_{max}R}{V_{th}}=\frac{1.35}{0.76}=1.78
\label{alpha}
\end{equation}  
where the voltage $i_{max}R$ is $(3 mA)(0.45 k\Omega) = 1.35 V$ and $V_{th}=0.76V$.  Protein decay rates are $\beta_1=\frac{R_y}{R_x}=\frac{470}{520}=0.90$, and $\beta_2=1$.  The Hill coefficient $n$ comes from
\begin{equation}
\label{n_alpha}
n\alpha=1.17G_1G_{-2}
\end{equation}  
where $G_1=-1.1$ and $G_{-2}=-3.3$ are closed loop gains of U1 and U2 in Fig.\ \ref{fig:gene} resulting in $n=(1.17)(1.1)(3.3)/1.78=2.4$. The characteristic time is $R_yC = (470\Omega)(2\times 10^{-8}f) = 9.4 \mu s$, so the dimensionless time is $t^{\prime} =\frac {t}{R_yC}$.

The high value $I_{ON}$ of the encoding signals $I_1,I_2$ is given by
\begin{equation}
I_{ON} = \frac{R_y\times(1Volt)}{R_IV_{th}}=\frac{470}{R_I(0.76)}=\frac{618}{R_I}
\label{ion}
\end{equation}
where $I_{ON}$ is changed by varying $R_I$ in Fig.\ \ref{fig:lsr_network}, i.e. $R_{I}=10k\Omega$ gives $I_{ON}=0.062$. A non-zero value of $g_2$ in Eq.\ 3 is achieved in the circuit by including a third current-sourcing transistor in Fig.\ \ref{fig:lsr_network} in the same way as the two
encoding signal transistors, but with the emitter resistor labelled $R_g$ and the input grounded so that the transistor provides a constant current $i_g=1Volt/R_g$. $g_2$ is changed by varying $R_g$ in the same way $R_I$ controls $I_{ON}$. The non-zero $g_2$ adds a term $R_yi_g$ to Eq.\ 5, where
\begin{equation}
g_2=\frac{R_yi_g}{V_{th}}=\frac{470}{R_g0.76}=\frac{618}{R_g}.
\label{i_g}
\end{equation}
$R_g=10k\Omega$ gives $g_2=0.062$.  

The $V_{noise}$ terms in Eqs.\ \eqref{circ_eq1}-\eqref{circ_eq2} approximate white noise voltages. Each $V_{noise}$ is characterized by its measured $rms$ value $V_{Nrms}$ and bandwidth. $V_{Nrms}$ is controlled by changing the gain via the potentiometer in Fig.\ \ref{fig:noise_circuit}. Noise strength $D$ in Eqs.\ \eqref{gene1}-\eqref{gene2} is given by 
\begin{equation}
D=\frac{V_{Nrms}}{V_{th}\sqrt{\gamma\,RC\,f_{c1}}}=\frac{V_{Nrms}}{0.76\sqrt{\gamma(9.4\mu s)(1.5 MHz)}}
\label{noise_strength}
\end{equation}
where $f_{c1}=1.5$ MHz is the cut-off frequency for the amplifier in Fig.\ \ref{fig:noise_circuit} and $\gamma$ decreases from $\pi/2$ at low gains to $\pi/4$ at high gain (when the potentiometer is set to $20k\Omega$ in Fig.\ \ref{fig:noise_circuit}). 

\subsection*{Circuit Analysis and Simulation}
Here we describe the circuit analysis and derive the connections between the model parameters used in Eqs.\ \eqref{gene1}-\eqref{gene2} and the circuit parameters. Further details are given in Refs.\ \cite{ed,ed2}.

The single gene circuit in Fig.\ \ref{fig:gene} is designed to reproduce the Hill function inhibition in Eqs.\ \eqref{gene1}-\eqref{gene2}. The op-amp U1 is configured as a subtraction amplifier with gain $G_1=-\frac{11}{10}=-1.1$. Replication of the Hill function behavior is achieved by allowing saturation of the output of the op-amp U2 and by having different unsaturated gains $G_{+2}$ and $G_{-2}$ for $V_{in}>V_{cth}$ and $V_{in}<V_{cth}$, respectively, due to the diodes in the feedback for U2. $G_{-2}$ is the gain of U2 when its output goes negative, in which case the diodes are not conducting, and therefore $G_{-2}=-\frac{3.3}{1.0}=-3.3$. $G_{+2}$ is a diminishing gain when the output of U2 becomes increasingly positive causing the diodes to go into conduction.  An increasing repressor concentration corresponds to $V_{in}$ surpassing $V_{cth}$ which causes the unsaturated output at U2 to change from a negative voltage of $G_1G_{-2}(V_{in}-V_{cth})$ to a positive voltage $G_1G_{+2}(V_{in}-V_{cth})$. The increasing voltage at the output of U2 turns the transistor off ($i_t \rightarrow 0$) which corresponds to complete inhibition of protein expression. Maximal protein expression $\alpha_{1,2}$ in Eqs.\ \eqref{gene1}-\eqref{gene2} corresponds to the maximum value of $i_t$, designated $i_{max}$.  $i_t=i_{max}$ occurs when $V_{in}=0$ (no repressor) because the output of U2 saturates at $V_{-sat}=-3.5$ V (for the LF412 op-amp supplied by $\pm 5$ V), resulting in a 0.65 V drop across the $220\Omega$ and therefore $i_{max}=3$ mA.

Circuit parameters for $\alpha$ and $\beta$ are found by using $V_{th}$ to convert Eqs.\ \eqref{circ_eq1}-\eqref{circ_eq2} to a dimensionless form for comparison to Eqs.\ \eqref{gene1}-\eqref{gene2}. The $0.45 k\Omega$ used for $R$ in Eq.\ \eqref{alpha} comes from the $R_y = 470 \Omega$ being nearly in parallel with the resistance (10 $k\Omega$) at the input to the subtraction amplifier U1 for gene-$x$. The relation for Hill coefficient $n$ is found by adjusting the dependence of $i_t$ on $V_{in}$ to match the slope of the Hill function $1/(1+x^n)$ at $x=1$ resulting in the relation \cite{ed2}:
\begin{equation}
\frac{-fV_{th}G_1G_{-2}}{f(5-V_{-sat})-0.6}=\frac{-n}{4}.
\label{slope} 
\end{equation} 
In Fig.\ \ref{fig:gene} the voltage divider fraction $f=0.4/2.6=0.154$ and $V_{-sat}=-3.5$ V. Using Eq.\ \eqref{alpha} in Eq.\ \eqref{slope} yields Eq.\ \eqref{n_alpha}. 

Parameter $V_{th}$'s correspondence to binding affinity of repressor to DNA is seen by noting that Eqs.\ \eqref{gene1}-\eqref{gene2} are dimensionless, meaning that in the process of going from chemical kinetic equations to Eqs.\ \eqref{gene1}-\eqref{gene2} the maximal expression rate $\alpha$ has been scaled by the repressor's inhibition binding constant $K_i$, and by a mRNA degradation rate \cite{elowitz2,ed2}. From Eq.\ \eqref{alpha} it follows that $V_{th}$ is proportional to $K_i$ since $\alpha$ is inversely  proportional to $K_i$ due to the scaling. To find the relation between $V_{cth}$ in Fig.\ \ref{fig:gene} and $V_{th}$ we note that the Hill function equals 0.5 when $x=1$.  Therefore $i_t$ must be half its maximum value when $V_{in}=V_{th}$ which gives \cite{ed2} 
\begin{equation*}
\frac{i_t}{i_{max}}=\frac{f(5-G_1G_{-2}(V_{th}-V_{cth}))-0.55}{f(5-V_{-sat})-0.6}=0.5.
\end{equation*}
Solving gives $V_{cth}\approx V_{th}+\frac{1}{G_1G_{-2}}$. Using $n=2.4$ and $\alpha = 1.78$ in Eqs.\ \eqref{alpha}-\eqref{n_alpha} gives: $G_1G_{-2}=3.65$, satisfied by $G_1=-1.1$ and $G_{-2}=-3.3$; $V_{th}=1.35/1.78 = 0.76$ V; and $V_{cth}=0.76 + 1/3.65 = 1.03$ V.

Comparing Eqs.\ \eqref{gene2} and \eqref{circ_eq2} shows that the encoding signals $I_1,I_2$ are related to the transistor currents  $i_1,i_2$ by:
\begin{equation}
I_{1,2} = \frac{R_yi_{1,2}}{V_{th}} .
\label{encoding}
\end{equation} 
The encoding currents $i_{1,2}$ take on two possible values depending on whether their logic train input in Fig.\ \ref{fig:lsr_network} is high or low. When the input is high ($> 4$V) the transistor is off so the current is zero. When the input is zero, the voltage divider consisting of the $4.7k\Omega$ and $2.2k\Omega$ produces 1Volt across the $R_I$ connected to the emitter of the \textit{pnp} transistor creating current $i_{1,2} = (1 Volt)/R_I$. Equation \eqref{encoding} then gives Eq.\ \eqref{ion} for $I_{ON}$.  Results of an analysis for a non-zero value of parameter $g_2$ are the same as for encoding signals $I_1,I_2$ and currents $i_1,i_2$ because $g_2$'s sourcing transistor is set up in the same way as the transistors in Fig.\ \ref{fig:lsr_network} for the encoding signals. Thus the non-zero value of $g_2$ is Eq.\ \eqref{i_g}.

Here we show how to use simulations to predict the circuit results. In the process we find Eq.\ \eqref{noise_strength}, the connection between circuit parameters and the noise strength $D$ in Eqs.\ \eqref{gene1}-\eqref{gene2}. A standard Euler-Maruyama simulation of Eq.\ \eqref{gene1} is
\begin{equation}
x_{i+1}-x_{i}=\left(\frac{\alpha_1}{1+y_i^{n}}-\beta_1 x_i + g_1\right)\Delta t^\prime + D\sqrt{\Delta t^\prime}\, \eta_i\left(0,1\right)
\label{E-M-1}
\end{equation} 
where $\eta _i(0,1)$ is a unit variance zero mean normal random distribution. The noise circuit in Fig.\ \ref{fig:noise_circuit} produces a measurable \textit{rms} voltage $V_{Nrms}$ consisting of contributions from all the frequency components present in the noise. The variance of the noise is the integral of its spectral density function $SD(f)$ over frequency, and is obtained from a measurement of $V_{Nrms}$;
\begin{equation}
V_{Nrms}^2 = \int{SD(f)\, df}.
\label{int_sdf}
\end{equation}
Idealized white noise assumed in Eqs.\ \eqref{gene1}-\eqref{gene2} has a $SD(f)$ which is uniform over an infinite bandwidth. However for the real noise from the 2-stage noise amplifier circuit in Fig.\ \ref{fig:noise_circuit} each op-amp's gain-bandwidth product produces a high frequency cut-off, $f_{c1}$ and $f_{c2}$. The resulting $SD(f)$ has the form
\begin{equation}
SD(f)=\frac{SD_0}{\left(1+\left(f/f_{c1}\right)^2\right)\left(1+\left(f/f_{c2}\right)^2\right)}
\label{sdf}
\end{equation} 
where $SD_0$ is a constant related to the strength of the noise.  The OPA228 op-amp has a gain-bandwidth product 33 MHz, therefore the first stage in Fig.\ \ref{fig:noise_circuit} with fixed gain $22\times$ has cut-off, $f_{c1}=33/22 = 1.5$ MHz.  The second stage's cut-off $f_{c2}$ varies from $33$ to $1.5$ MHz depending on the gain setting determined by the potentiometer in the feedback of the second stage op-amp.  

The integration in Eqs.\ \eqref{int_sdf}-\eqref{sdf} gives 
\begin{equation}
V_{Nrms}^2 =\int_0^\infty \frac{SD_0}{\left(1+\left(f/f_{c1}\right)^2\right)\left(1+\left(f/f_{c2}\right)^2\right)}\,df= \frac{f_{c1}f_{c2}}{f_{c1}+f_{c2}}\frac{\pi}{2}\,SD_0.
\label{sd_int}
\end{equation}
There are two limiting cases for the integral: for small gains $f_{c2}\approx 33$ MHz $>>f_{c1}$ giving $(\pi/2)f_{c1}SD_0$; and for large gain (potentiometer $\rightarrow 20k\Omega$ in Fig.\ \ref{fig:noise_circuit}) $f_{c2}\approx f_{c1}$ giving $(\pi/4)f_{c1}SD_0$. Thus the integral in Eq.\ \eqref{sd_int} is $\gamma f_{c1}SD_0$ where $f_{c1}=1.5$ MHz and $\gamma$ varies from $\pi/2$ for small noise to $\pi/4$ for large noise. For frequencies within the noise bandwidth ($<1.5$ MHz) the amplitude spectral density $ASD$ (units Volt Hz\textsuperscript{-1/2}) has a constant value given by
\begin{equation}
ASD_0=\sqrt{SD_0}=\frac{V_{Nrms}}{\sqrt{\gamma f_{c1}}}
\label{asd}
\end{equation}
$ASD_0$ is a good approximation of the white noise strength provided that the Fig.\ \ref{fig:lsr_network} genetic circuit's characteristic response rate $1/(RC)$ is much less than the noise bandwidth, meaning that $1/(RC) << 2\pi f_{c1}$. This condition is ensured since the  $RC$ used here is $(470\Omega)(0.02 \mu f)=9.4 \mu s$, and the bandwidth of the noise is $f_{c1}\approx 1.5$ MHz.

\begin{figure}[!ht]
\begin{center}
\includegraphics[width=3.2in]{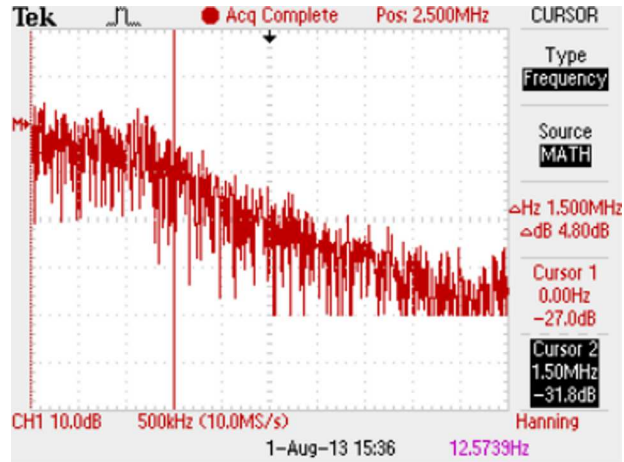}
\end{center}
\caption{{\bf Measured frequency spectrum of noise.} For 2-stage noise amplifier shown in Fig.\ \ref{fig:noise_circuit} with second stage gain $11\times$. Horizontal scale is 500 kHz/Div, so cursor indicates 1.5 MHz as location of cut-off frequency. From FFT function on Tektronix TDS 2024B oscilloscope. }
\label{fig:fft}  
\end{figure}

Figure \ref{fig:fft} shows the measured frequency content from the noise circuit when the potentiometer at the second stage is set for gain $11\times$ producing $V_{Nrms}=0.90$ V and $f_{c2}=33/11=3$ MHz.  In this case $f_{c2}=2f_{c1}$ and $\gamma=\pi/3$.  Figure \ref{fig:fft} shows that the frequency content is relatively flat out to the cut-off near 1.5 MHz and therefore is a good approximation to white noise for the genetic network circuit. $V_{Nrms}=0.90$ V is on the high end of the noise amplitudes used here. 

The circuit Eqs.\ \eqref{circ_eq1}-\eqref{circ_eq2} which need to be simulated are of the form  
\begin{equation}
\frac{dV}{dt}=\frac{f(V)}{RC}+\frac{V_{noise}}{RC}
\label{general}
\end{equation}
where $V_{noise}$ approximates a white noise voltage. $V_{noise}$ is characterized by its measured $rms$ value $V_{Nrms}$ and bandwidth, and Eq.\ \eqref{asd} gives the noise's amplitude spectral density. The Euler-Maruyama simulation for Eq.\ \ref{general} is 
\begin{equation}
V_{i+1}-V_i = \frac{f(V_i)}{RC}\Delta t +\frac{ASD_0}{RC}\sqrt{\Delta t}\,\eta_i\left(0,1\right).
\end{equation}
Using dimensionless time $\Delta t^\prime = \Delta t/(RC)$ and the measured noise amplitude $V_{Nrms}$ gives
\begin{equation}
V_{i+1}-V_i=f(V_i)\Delta t^\prime + \frac{V_{Nrms}}{\sqrt{\gamma\,RC\,f_{c1}}}\sqrt{\Delta t^\prime}\,\eta_i\left(0,1\right)
\label{E-M-3}
\end{equation}  
Normalizing by the voltage scale $V_{th}$ puts Eq.\ \eqref{E-M-3} in the form of Eq.\ \eqref{E-M-1} and gives the connection between circuit parameters and dimensionless noise amplitude $D$ shown in Eq.\ \eqref{noise_strength}. For example, using $V_{th}=0.76 V$, $RC=(470\Omega)(0.02 \mu f)$, $\gamma=\pi /2$, and $f_{c1}= 1.5$ MHz gives $D=0.28V_{Nrms}$. 

\section*{Results and Discussion}
\begin{figure}[!ht]
\begin{center}
\includegraphics[width=2.5in]{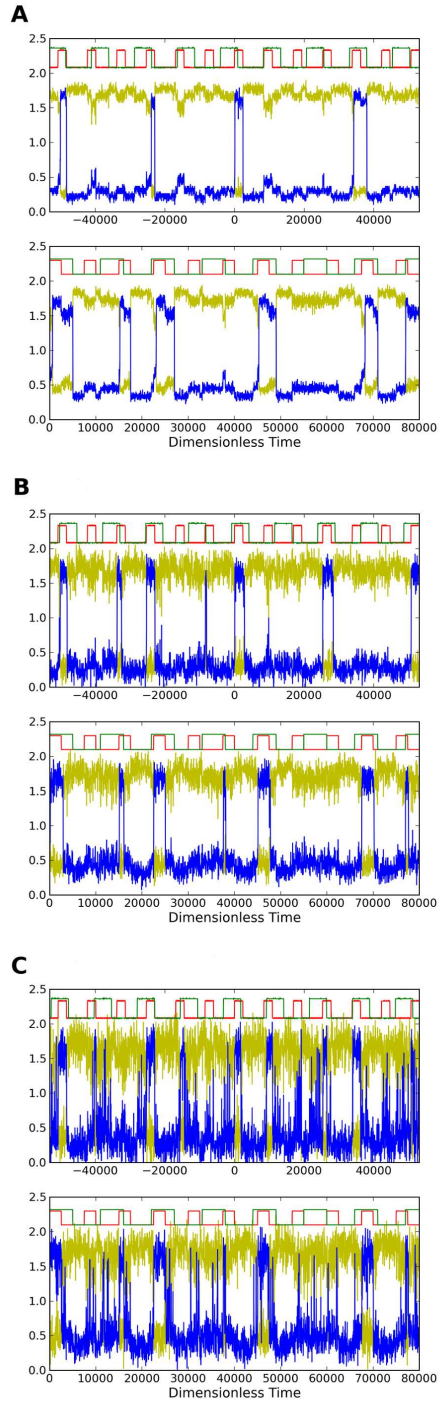} 
\end{center}    
\caption{{\bf Time series for circuit measurements (upper graph) and simulations (lower) for different noise strengths showing AND/NAND LSR.} Circuit shown in  Fig.\ \ref{fig:lsr_network}.  Simulations are of Eqns.\ \eqref{gene1}-\eqref{gene2}. Upper red and green traces indicate logic states of the encoding inputs $I_{1,2}$ and lower traces show the complementary outputs $x$(yellow) and $y$(blue). Panel (b) has noise level within the optimal range for displaying AND/NAND characteristics. Noise strengths in simulation and in circuit are: (a) $D=0.043$ and $V_{Nrms}=200mV$, (b) $D=0.107$ and $V_{Nrms}=500mV$, (c) $D=0.15$ and $V_{Nrms}=700mV$. Voltages and times for the circuit measurements have been converted to dimensionless quantities as described in the text.}
\label{fig:time_series}
\end{figure}

Figure \ref{fig:time_series} shows simulations and circuit measurements for three values of noise using parameters $n=2.4$, $\alpha=1.78$, $\beta_1=0.90$, $\beta_2=1$, $I_{ON}=0.067$, and $g_1=g_2=0$.  It is apparent that for an optimal noise level (Fig.\ \ref{fig:time_series}b) the circuit indeed performs the
logic AND/NAND function, and that for the smaller (Fig.\ \ref{fig:time_series}a) and larger (Fig.\ \ref{fig:time_series}c) noise values faithful logic response is lost. At the low noise (Fig.\ \ref{fig:time_series}a) the outputs sometimes fail to respond to the 0 to 1 transition from the AND of $I_1,I_2$, and for the 1 to 0 transition the outputs often wait until \textit{both} inputs go low before responding thereby causing a delayed response. In Fig.\ \ref{fig:time_series}b the responses are quick for both the up and down transition. Figure \ref{fig:time_series}c shows that at the high noise level the responses are again quick as in \ref{fig:time_series}b, but the outputs also make erroneous transitions.  The circuit behavior is seen to be in agreement with the simulations of Eqs.\ \eqref{gene1}-\eqref{gene2}. 

\begin{figure}[!ht]
\begin{center}
\includegraphics[width=6.2in]{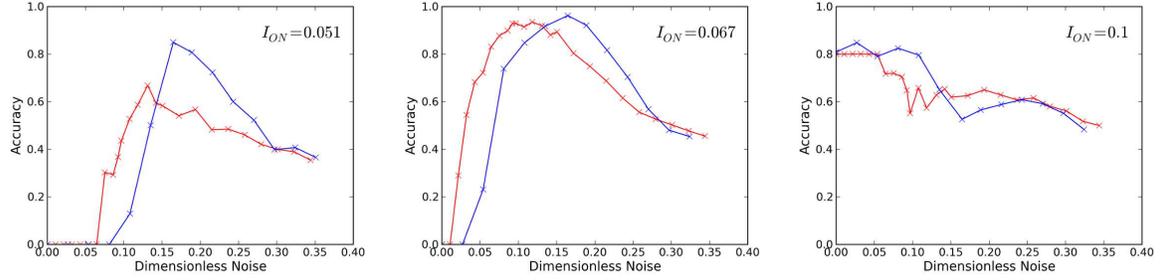}
\end{center}
\caption{{\bf Accuracy $a$ of the AND/NAND logic response for simulations (red) and circuit measurements (blue) as a function of noise strength $D$ and encoding amplitude $I_{ON}$.} Noise strength $V_{Nrms}$ for the circuit measurements has been converted to dimensionless strength $D$ using Eq.\ \eqref{noise_strength} with $\gamma=\pi/2$. }
\label{fig:accuracy}
\end{figure}

In order to investigate the range of noise strengths which produce accurate logic response, and to find optimal values for the amplitude $I_{ON}$ of the small signal inputs $I_{1,2}$ we  define an accuracy measure $a$,
\begin{equation}
a=a_L\times a_H
\end{equation}   
where $a_L$ is the percent of time the $x,y$ outputs are correct when the AND operation of $I_1$ and $I_2$ is low, and $a_H$ is the percentage of time correct when the AND operation is high. This definition has the property that when the $x,y$ outputs do not respond at all, then $a=0$ since $a_H=0$ even though $a_L=1$. The expectation then is that for no noise there should be no stochastic resonance response so that $a=0$, and that for extreme noise each accuracy would approach 50\% so that $a \rightarrow 0.25$. If $x,y$ respond immediately with no mistakes then $a=1$. Figure \ref{fig:accuracy} shows accuracy $a$ for simulations and circuit measurements as a function of noise strength for different values of encoding amplitude $I_{ON}$. 

Figure \ref{fig:accuracy}a shows that for a small value of encoding amplitude, $I_{ON}=0.051$ the network is not able to give a faithful logic response at any noise level.  The response at low noise and at high noise are as predicted, $a=0$ and $a \rightarrow 0.25$, respectively, but the peak of the window of stochastic resonance response is well below $1$. Figure \ref{fig:accuracy}b shows a window of noise providing faithful response for $I_{ON}=0.067$. The reason that the accuracy $a$ is slightly below 1 in the window is that the $x,y$ outputs do not respond immediately to the AND/NAND transitions. This time lag causes the percent of time with incorrect response to be non-zero and therefore $a_H$ and $a_L$ are less than 1. In principle an allowance for a time lag could be included in the calculation of $a$ if it were deemed necessary.  However, such an added complication would not make the noise window any more apparent than it already is in Fig.\ \ref{fig:accuracy}. Figure \ref{fig:accuracy}c shows that at a high value, $I_{ON}=0.10$, the outputs respond even with no noise, and the addition of noise only creates more errors.  The relative shift between the simulation and circuit accuracies is due to assumptions made about the noise spectral density function and the integration in Eq.\ \eqref{sd_int} leading to Eq.\ \eqref{E-M-3} which gives the connection between the measured noise amplitude $V_{Nrms}$ and dimensionless noise $D$. In the idealized case Eq.\ \eqref{sd_int} finds $\gamma$ ranges from $\pi/2$ to $\pi /4$ in Eq.\ \eqref{E-M-3}, with $\pi/2$ being appropriate for the noise levels used in Fig.\ \ref{fig:time_series}. Adjusting the value of $\gamma$ can eliminate the relative shift, but there is nothing to be gained since the appearance of an optimal noise window for LSR at an appropriate value of $I_{ON}$ is already apparent. 

\begin{figure}[!ht]
\begin{center}
\includegraphics[width=4in]{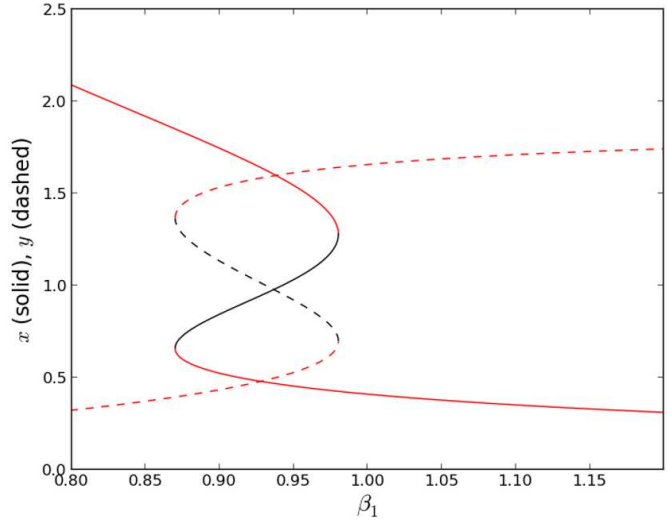}
\end{center}
\caption{{\bf Bifurcation diagram for state-variables $x$ (solid) and $y$ (dashed) configured for OR/NOR.} $x$ and $y$ are complementary outputs, when one is high the other is low. Red (black) indicate stable (unstable) fixed points. System is bistable for $0.870<\beta_1 <0.980$. Calculated for Eqs.\ \eqref{gene1}-\eqref{gene2} with $n=2.4$, $\alpha_i=1.78$, $\beta_2=1$, $g_1=0$, $g_2=0.062$, $I_i=0$, and $D=0$. }
\label{fig:beta_bif_g2}  
\end{figure}

\begin{figure}[!ht]
\begin{center}
\includegraphics[width=2.5in] {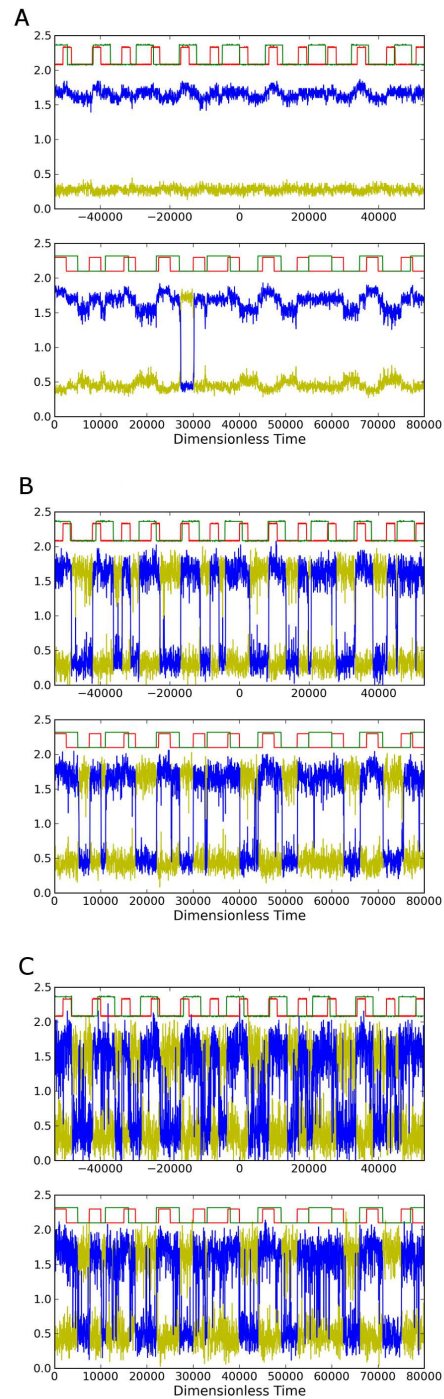}
\end{center}
\caption{{\bf Time series for circuit measurements (upper graph) and simulations (lower) for different noise strengths showing OR/NOR LSR. } Panel (b) has noise level within the optimal range for displaying OR/NOR characteristics. Noise strengths in simulation and in circuit are: (a) $D=0.043$ and $V_{Nrms}=200mV$, (b) $D=0.107$ and $V_{Nrms}=500mV$, (c) $D=0.15$ and $V_{Nrms}=700mV$. Voltages and times for the circuit measurements have been converted to dimensionless quantities as described in the text.}
\label{fig:lsr_ornor}
\end{figure}

One can also {\em reconfigure} the system to another set of logic
functions, namely the fundamental OR/NOR logic, by simply including a non-zero value for $g_2$. The parameter $g_2$ effectively changes
the relative position and depth of the wells of the bistable system, 
allowing the response to morph from AND/NAND to
OR/NOR. For instance changing $g_2$ from $0$ to $0.062$ (with all other
parameters unchanged) changes the bifurcation diagram from that in Fig.\ \ref{fig:beta_bif} to Fig.\ \ref{fig:beta_bif_g2} showing that the system is now bistable for $\beta_1=0.9$. The result is that the system displays a clear OR and the complementary NOR
response as shown by the simulation and circuit results in Fig.\ \ref{fig:lsr_ornor}. The low noise case Fig.\ \ref{fig:lsr_ornor}a shows that at this low noise level the system usually fails to respond.  Figure \ref{fig:lsr_ornor}a also shows the resting states are reversed from the $g_2=0$ case in Fig.\ \ref{fig:time_series}a. Figure \ref{fig:lsr_ornor}b shows the OR/NOR response at a noise value within the window for LSR, and Fig.\ \ref{fig:time_series}c shows errors when the noise is too large.  

In summary, our results show that the dynamics of the two variables
$x$ and $y$ with $g_2 = 0$, mirror AND and the complementary NAND
gate characteristics. Further, when $g_2 \ne 0$, we obtain a clearly
defined OR/NOR gate.  Since $x$ is low when $y$ is high and
vice-versa, the dynamics of the two variables always yield {\em
complementary} logical outputs, simultaneously. That is, if $x(t)$
operates as NAND/NOR, $y(t)$ will give AND/OR. 

These results extend the scope and indicate the generality of
the recently observed phenomena of Logical Stochastic Resonance
through experimental verifications. Further, these observations may
provide an understanding of the information processing capacity of
synthetic genetic networks, with noise aiding logic patterns. It also
may have potential applications in the design of biologically inspired
gates with added capacity of reconfigurability of logic operations.  

We have also demonstrated that the electronic circuit provides an additional tool for investigating dynamics of proposed genetic networks. The circuit measurements are complementary to numerical simulations, thereby giving indication of the robustness of a particular network design and potential for successful realization in a biological system.   

Thus the results presented in this work suggest new directions in
biomolecular computing, and indicate how robust computation may be
occurring at the scale of regulatory and signalling pathways in
individual cells. Design and
engineering of such biologically inspired computing systems not only
present new paradigms of computation, but can also potentially enhance
our ability to study and control biological systems \cite{nature}.

\section*{Acknowledgments}
S.K.D. acknowledges support by the CSIR Network project GENESIS (BC0123) and the CSIR Emeritus Scientist scheme.


\bibliography{stoc_res,lsr_hellen}

\section*{Figure Legends}

\section*{Tables}
\begin{table}[!ht]
\caption{
\bf{Logic Table}}
\begin{tabular}{|c|c|c|c|c|}
\hline
Input Set ($I_1$,$I_2$)&OR&AND&NOR&NAND\\ \hline
\hline
(0,0) &0&0&1&1\\ 
(0,1)/(1,0) &1&0&0&1\\ 
(1,1) &1&1&0&0\\ \hline
\end{tabular}
\begin{flushleft}Relationship between the two inputs and the output of the
 fundamental OR, AND, NOR and NAND logic operations. Note
  that the four distinct possible input sets $(0,0)$, $(0,1)$, $(1,0)$
  and $(1,1)$ reduce to three conditions as $(0,1)$ and $(1,0)$ are
  symmetric. Note that {\em any} logical circuit can be constructed by
  combining the NOR (or the NAND) gates \cite{mano1,bartee}.
\end{flushleft}
\label{table1}
\end{table}

\end{document}